\documentclass[12pt]{article}

\usepackage{scrextend} %Allows the addmargin environment, for greater control over the margins of a section
\usepackage[margin=1in]{geometry}
\usepackage{amsmath,amsthm,amssymb}
\usepackage{mdwlist}%Adds the \suspend and \resume commands for environments (or at leasy enumerations)
\usepackage{graphicx}
\usepackage{caption}
\usepackage{subcaption}
\usepackage{float}
\usepackage{pdfpages}

\newcommand{\Matlab}{\textsc{Matlab}}

\title{Numerical Approximation of Kramers-Kronig\\Relations to Transform Discretized Absorption Data}
\author{Patrick D Fitzgerald}
\date{December 20, 2017} % Hard coded to match original document in perpetuity

\begin{document}

\maketitle
\begin{abstract}
    The Kramers-Kronig relations describe a pair of integral transforms relating the real and imaginary components of an analytic function in the complex plane. These relations are particularly useful in extracting refractive index characteristics of a given physical test species, wherein more direct measurements are not terribly reliable. In this report, a method of performing this integral transformation is derived and discussed. Special attention is given to the precision of these methodologies, and a specific implementation of this in \Matlab{} is presented.
\end{abstract}

\section{Introduction}
    In the field of spectroscopy, ascertaining material properties such as absorption coefficient and refractive index is frequently of interest in order to characterize them, and one way to measure these properties is through directly passing an electromagnetic pulse through a sample, and observing how the waveform is distorted and attenuated. Due to the nature of this type of measurement scheme, called Time Domain Spectroscopy, the resulting absorption profile is generated with higher certainty than that of the refractive index. This presents an opportunity to apply some interesting underlying physics and mathematics to transfer some of that certainty over to the less certain quantities.
    
    Resulting from the type of measurement, it is most straightforward to calculate the absorption coefficient directly for a uniform range of frequencies, and to computationally generate the refractive index from the absorption data. 

\subsection{Absorption and Refractive Index}
    Generally speaking, the absorption coefficient and the refractive index are two electromagnetic properties which help describe how certain materials behave under the effect of excitation by light. Further implications regarding the microscopic behaviors and phenomena of the material can be derived from these quantities. The absorption coefficient, $\alpha$, represents the proportion of the light passing through a material which it absorbs per unit length. The refractive index, $n$, determines the amount that light slows down within the medium. Both of these quantities depend on the frequency of the light in question, and as such they can be seen as functions of the angular frequency $\omega$ (rad/s) or (ordinary) frequency $\nu$ (Hz), where $\omega=2\pi\nu$. Interestingly, these quantities are strongly related, and have a more obvious representation as the dielectric function
    \begin{equation}
        \varepsilon(\omega) = n(\omega) + i\frac{c}{2\omega}\alpha(\omega),
    \end{equation}
    where $c$ is the speed of light in vacuum. Due to physical and temporal symmetry considerations, this function is assumed to have the property that
    \begin{equation}
        \varepsilon(-\omega) = \varepsilon(\omega)^*
    \end{equation}
    where the asterisk designates the complex conjugate. Accordingly, the refractive index and the absorption coefficient become even functions about the origin. $\varepsilon$ is an analytic function in the upper half of the complex plane, which means that the results of Cauchy's complex contour integral theorems become applicable and exploitable.

\subsection{The Kramers-Kronig Relations and Derivatives}
    One result of those contour integral theorems is the following relations, which may be used to link together the real and imaginary components of an analytic function $f$:
    \begin{subequations}
    \begin{equation}
        \text{Re}(f(z)) = \frac{1}{\pi}\mathcal{P}\int_{-\infty}^{\infty}\frac{\text{Im}\big(f(z')\big)}{(z'-z)}dz' \label{eq2a}
    \end{equation}
    and
    \begin{equation}
        \text{Im}(f(z)) = -\frac{1}{\pi}\mathcal{P}\int_{-\infty}^{\infty}\frac{\text{Re}\big(f(z')\big)}{(z'-z)}dz'
    \end{equation}
    \end{subequations}
    where $z$ is a free parameter (fixed for any given integration), and $z'$ is a dummy integration variable. These relations are called the Kramers-Kronig relations, or the dispersion relations. It is clear to see that under most scenarios for $f$, there is a pole present at $z'=z$. The $\mathcal{P}$ represents the Cauchy Principal Value, which indicates that the integral must be computed carefully around the pole so as to have it converge on a nontrivial, nonarbitrary value (if it does converge at all). For example, one way of representing this careful integration for some general function in the numerator $g$ is
    \begin{equation}
        \mathcal{P}\int_{-\infty}^{\infty}\frac{g(z')}{z'-z}dz' = \lim_{\epsilon\rightarrow0} \bigg(\int_{-\infty}^{z-\epsilon}\frac{g(z')}{z'-z}dz' + \int_{z+\epsilon}^{\infty}\frac{g(z')}{z'-z}dz'\bigg),
    \end{equation}
    which just means that the integration approaches the pole from either side at equal rates.\footnote{If this choice were not taken, and the pole was integrated closer to the poles at different rates from the left and the right, then the whole integral could result in any arbitrary value, so long as each side was divergent. This choice of equal rates corresponds directly to the area-under-the-curve perspective of integrals, whereby an equal amount of infinite area above and below the axis logically allows for cancellation, yielding a finite answer. This care must be taken for each pole present in the integrand.}$^,$\footnote{In addition to specifying equal rates when approaching a singularity, this prescription also similarly describes the rates at which one must approach positive and negative infinity. This is not of concern, since the functions being integrated in this context are assumed to decay very quickly at higher frequencies, and thus converge to the same value regardless of the rates of approach.}

    By applying this to the present context, $f$ may be replaced with $\varepsilon$. From the resulting relation corresponding to (\ref{eq2a}), we can generate two integral relations, one where the free parameter is relabeled to $\omega$, the other where the free parameter is relabeled to $\omega_a$, and in both the dummy integration variable is relabeled to $\omega'$. These two relations may be combined by taking their difference and combining the integrals into one, whose bounds may be halved due to the symmetry present in $\alpha$, with slight modification to the integrand of each. This combination and simplification leaves the Singly Subtractive Kramers-Kronig Relations (SSKKR)\cite{coutaz}:
    \begin{equation}
        n(\omega) = n(\omega_a) + \frac{c}{\pi}\mathcal{P}\int_0^{\infty} \frac{\alpha(\omega')(\omega^2-\omega_a^2)}{(\omega'^2-\omega^2)(\omega'^2-\omega_a^2)}d\omega'.\label{eq5}
    \end{equation}
    This result allows for the determination of the refractive index at an arbitrary $\omega$ from information derived from the absorption coefficient over a large region, and yielded relative to the refractive index at the frequency $\omega_a$, appropriately named the anchor frequency. The intent of this anchor frequency is to select a frequency at which the refractive index is known with decent certainty, so that the rest of the desired refractive index values may be defined relative to this reference. Another benefit of this form is the relative size of the factor which scales $\alpha$ in the integrand: because it decays rapidly away from respective poles, it reduces the total contribution of the part of the integral which covers the arbitrarily high $\omega'$ values, or equivalently values of $\omega'$ close to zero. This means that truncating the integral's bounds produces less total error. This is especially useful in mitigating the effects of a finite bandwidth of data for $\alpha$.

\section{Numerically Approximating the SSKKR}
    Due to the fact that values for the absorption coefficient have only been measured as a discrete set of data at equispaced frequency values, the integral in (\ref{eq5}) must be approximated by a discrete summation of weighted absorption values of the form
    \begin{equation}
        n_j = n(\omega_a) + \sum_{i=1}^N{}w_i^{(j)}\alpha_i,
    \end{equation}
    where $N$ is the total number of data points available for $\alpha$, and the subscripts on $n_j$ and $\alpha_i$ are the refractive index and absorption coefficient corresponding to frequencies $\omega_j$ and $\omega_i$, respectively (or equivalently $\nu_j$ and $\nu_i$, respectively). $w_i^{(j)}$ serves as a weighting coefficient for each of the measured values of $\alpha$, and will need to be chosen to best approximate the integral. The subscript on $w$ defines which $\alpha$ value it should scale (corresponding to $\omega'$ in (\ref{eq5})), and the superscript defines where the value for $n$ is being calculated (corresponding to $\omega$ in (\ref{eq5})). 
    
    Since the $\alpha$ data is equispaced, the available methods for standard numerical quadrature are restricted to schemes which support this structure, and the fact that no extra intermediate data points are available\footnote{
        While it is technically possible to generate more intermediate data points from the discrete Fourier transform from which all of this data was generated, this would be an overly complicated way to tackle this problem. Furthermore, choosing finer meshes is not a proper substitute for an intelligent management of the precision of the computation.}
    similarly restricts the available choices. Accordingly, Newton-Cotes (NC) quadrature schemes will primarily be used herein. Where these standard NC formulas are not used, new formulas will be derived, but the construction of these will involve the same logic from which these standard NC formulas were originally derived.
    
    Before delving into the details of the evaluation, let us first make an alteration of our perspective of this integral: it is more convenient to work with respect to $\nu$, since the corresponding numerical values for frequency are easier to interpret as ordinary frequency, instead of angular frequency\footnote{This integral transform will be used primarily on data generated from terahertz spectroscopy, and the frequencies listed in THz ($10^{12}$Hz) is often more convenient in this respect. To avoid issues of integrating over very large domains, the units for $\nu$ are just left in THz within the \Matlab{} code, and other physical constants like $c$ are converted to this same set of units.}:
    \begin{equation}
        n(\nu) = n(\nu_a) + \frac{c}{2\pi^2}\mathcal{P}\int_0^{\infty} \frac{\alpha(\nu')(\nu^2-\nu_a^2)}{(\nu'^2-\nu^2)(\nu'^2-\nu_a^2)}d\nu'.\label{eq7}
    \end{equation}
    No factor of $2\pi$ is generated from this change of variables inside $n$ or $\alpha$, since as far as these functions are concerned, their values are just labeled by the corresponding $\omega$ and $\nu$ values, but are not explicitly dependent on them, in a numerical sense.

    \begin{figure}[H]
        \centering
        \begin{minipage}[t]{0.48\textwidth}
            \centering
            \includegraphics[width=1.0\textwidth]{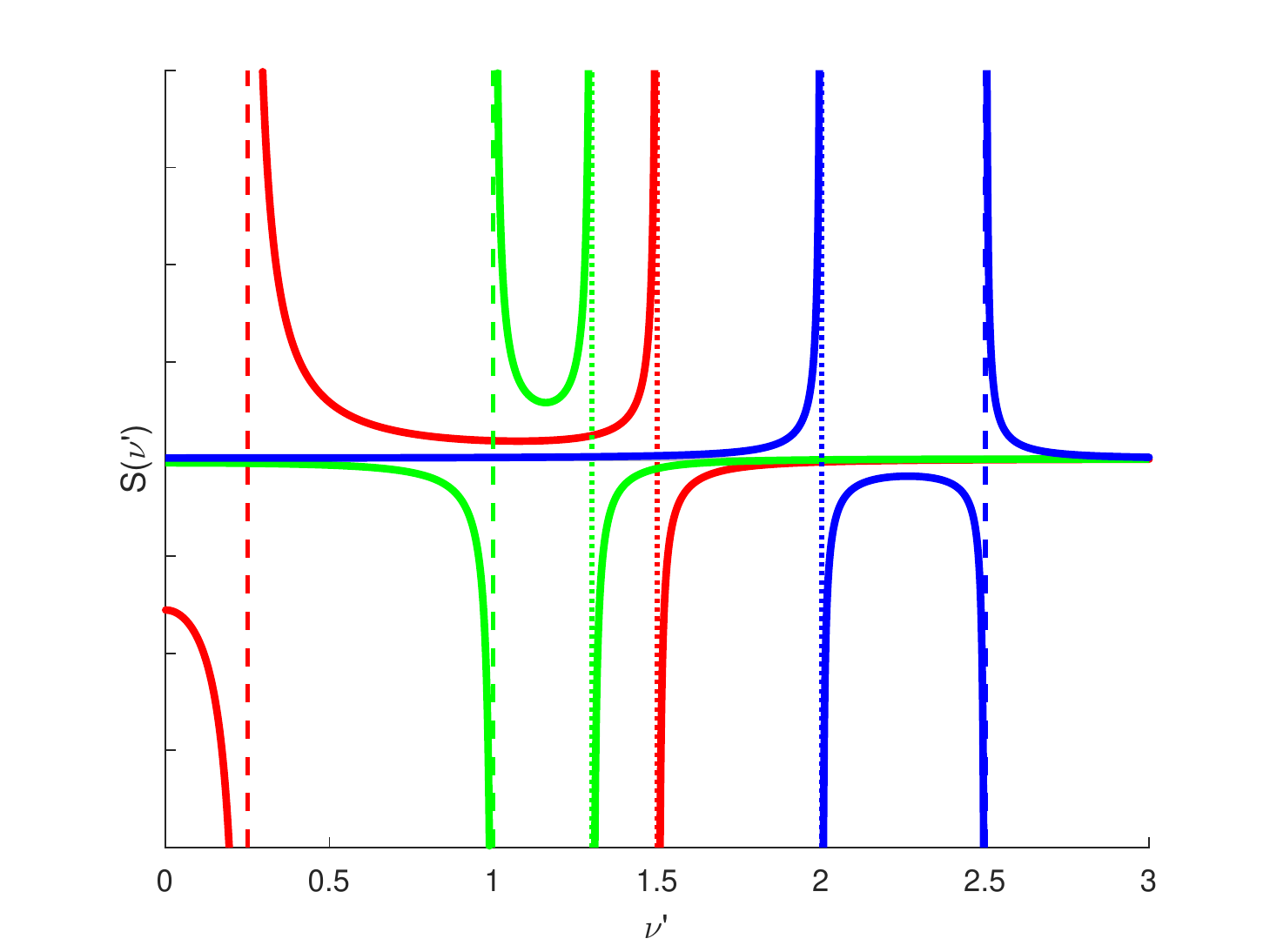}
            \caption{A sample of $S$ functions, where the $\nu$ and $\nu_a$ are placed at a variety of locations within the domain. The dashed vertical lines correspond to $\nu'=\nu$, and the dotted vertical lines correspond to $\nu'=\nu_a$.}\label{fig1}
        \end{minipage}%
        \begin{minipage}[t]{0.035\textwidth}
        \hspace{0.03\textwidth}
        \end{minipage}%
        \begin{minipage}[t]{0.48\textwidth}
            \centering
            \includegraphics[width=1.0\textwidth]{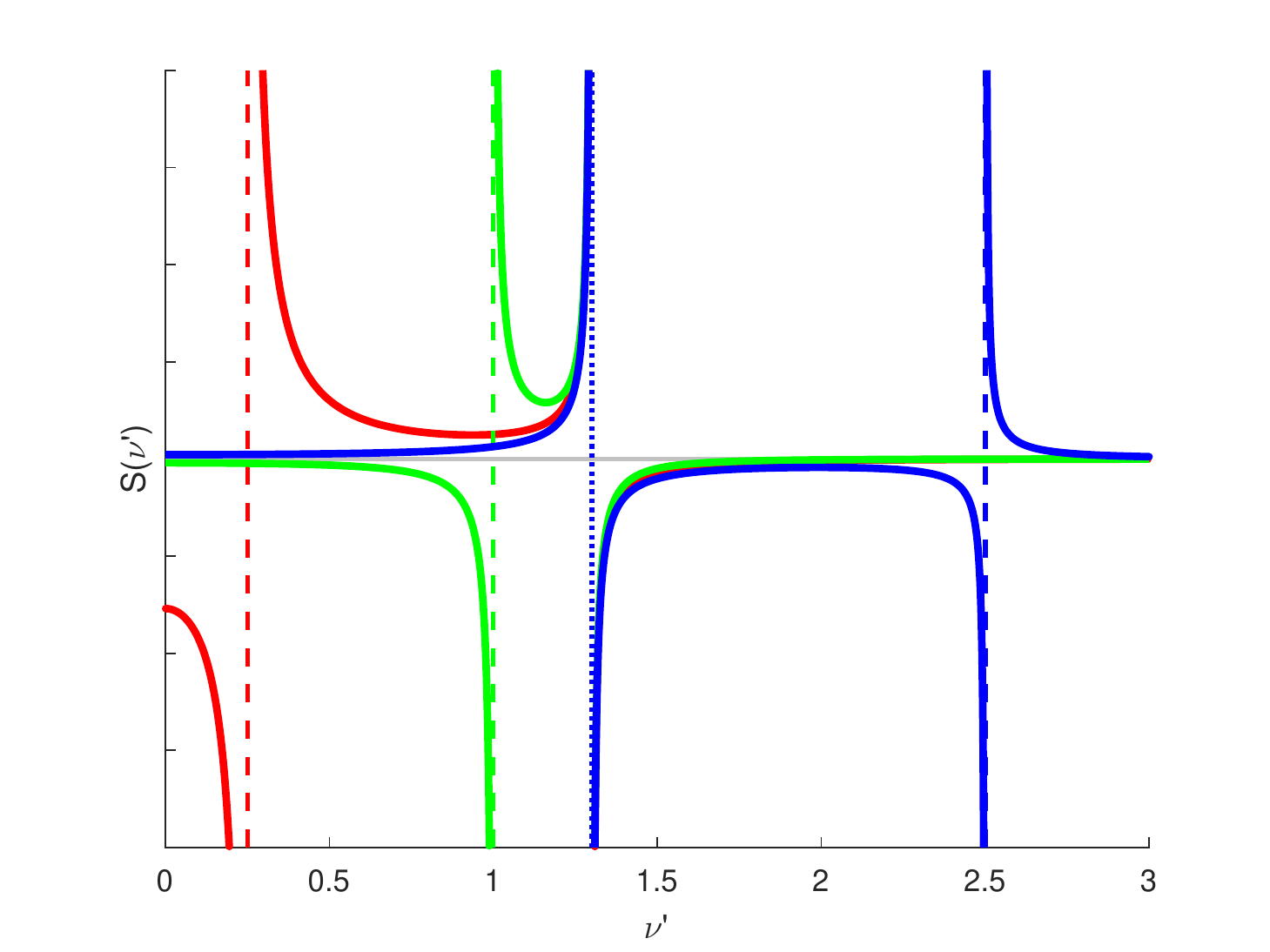}
            \caption{The same as in Figure \ref{fig1}, except that $\nu_a$ is fixed at $\nu'=1.3$, while $\nu$ alone is varied. This fixed nature of $\nu_a$ is properly representative of the different scenarios which are seen in this discussion.}\label{fig2}
        \end{minipage}
    \end{figure}

\subsection{Consideration of Different Cases}
    In attempting to approximate anything, it is often a good first step to appreciate what details are most important, so that they may be the primary focus of the approximation. After those are understood and accounted for, it is also important to understand any sort of cases which require specific behavior to maintain a high quality of approximation. 
    
    Regarding the former, in order to gain an appreciation of the important details of this integrand, we may simply focus on just the function which scales $\alpha$, which we may name $S(\nu')$, where the value for $\nu$ and $\nu_a$ are simply interpreted as parameters:
    \begin{equation}
        S(\nu') = \frac{(\nu^2-\nu_a^2)}{(\nu'^2-\nu^2)(\nu'^2-\nu_a^2)}
    \end{equation}
    A variety of scenarios for $S(\nu')$ are presented in Figures \ref{fig1} and \ref{fig2}, which each have different values for $\nu$ and $\nu_a$. From this graphical representation, it is clear to see that the region just next to the poles will contribute strongly to the integral and therefore also to the weights $w_i^{(j)}$. Outside these neighborhoods, the functions are relatively slow to change, and in many cases, rather small in magnitude. From these observations, it is reasonable to assume that more care needs to be given to the regions around the poles\footnote{
        In addition to needing care due to the Cauchy Principal Value, this region needs extra care due to its asymptotic behavior. Since all of the NC schemes assume some sort of terminating polynomial form in the domain of integration, they are especially poorly equipped to handle functions which are not well approximated by their lower degree terms, as is characteristic of asymptotic forms.
    }, and a less nuanced attack will suffice elsewhere.
    
    Since for a given set of computations for determining $n(\nu)$, the value for $\nu_a$ will remain fixed and the value for $\nu$ will iterate over a range of values, there are necessarily more cases than just 'near pole' and 'not near pole'. The 'near pole' case must be broken into three categories: only one pole nearby, two distinct poles nearby, and one second-degree pole nearby. In order to properly discuss these cases, let us label them as shown in Table \ref{labels}:
    
    \begin{table}[H]
        \centering
        \caption{Node proximity-to-poles labels}
        \label{labels}
        \begin{tabular}{r|l}
            Case  & Description              \\ \hline
            I     & Not near any poles       \\
            II    & Near one pole            \\
            III   & Near two distinct poles  \\
            IV    & Near one second degree pole
        \end{tabular}
    \end{table}

\subsection{Case I: Regions Away from Poles}
    Outside any asymptotic regions, we will simply apply standard composite NC schemes. Since this case represents all the locations where nothing especially drastic is happening, the regions to which this composite scheme will need to be applied might not always be of convenient or consistent size\footnote{This region essentially acts as a filler, in order to gather up the contribution of the $\alpha$ values not described elsewhere by more nuanced means.}.
    Accordingly, a plan needs to be in place on how to handle nonstandard size. If a minimum NC order has $N_{I,min}+1$ points, then order could be increased to represent more points ($N_{I}+1$) without an increase in truncation error. This would eventually create significant round-off error, so trying to increase the order of many segments by a small amount is better than increasing one drastically. If, however, this Case I region is too small due to being pinched between a region of Case II and that of another Case II or the edge of the available data, then the Case II region should be extended to fully represent that otherwise-Case-I region. A depiction of this Case II extension can be most easily seen at the top and bottom corners of Figure \ref{matrix}. 
    
    In these segments of Case I points, the weight values will contain the inherent NC weights, and the contribution of the integrand weighting function, $S(\nu')$, will manifest simply as scaling the NC weights by the $S$ value at the respective $\nu'$ values. Since these segments are far from poles, $S$ will be well behaved.

\subsection{Cases II-IV: Regions Next to Poles}
    Case II regions will require a method of approximating the respective integrals with a scheme with the same underlying structure as the NC scheme. If inside this region of $N_{II,min}+1$ points, and the corresponding integral over them, it becomes beneficial to find an interpolating function for $\alpha$ in this range. In order to do this, let us employ a Lagrange interpolating polynomial, in the same fashion as the NC scheme does. Thus,
    \begin{equation}
        \alpha(\nu')\approx\sum_{i=0}^{N_{II}}\bigg(\alpha_i\prod_{\substack{k=0\\k\neq{}i}}^{N_{II}}\frac{\nu'-\nu_k}{\nu_i-\nu_k}\bigg)
    \end{equation}
    in the region from $\nu_0$ to $\nu_{N_{II}}$ (which may easily be relabeled to represent any such integration without loss of generality). Since there is only one pole to be careful of, it would be most convenient to evaluate the rest in the previous manner. To this end, we may expand the integrand weighting function $S$ to be
    \begin{equation}
        S(\nu')=\frac{-1}{2\nu}\bigg(\frac{1}{\nu'+\nu}\bigg)+\frac{1}{2\nu}\bigg(\frac{1}{\nu'-\nu}\bigg)+\frac{-1}{2\nu_a}\bigg(\frac{1}{\nu'-\nu_a}\bigg)+\frac{1}{2\nu_a}\bigg(\frac{1}{\nu'+\nu_a}\bigg),
    \end{equation}
    by partial fraction decomposition. In this form, it becomes very easy to extract one of the pole's respective components and leave the rest to be evaluated as in the Case I scenario. For the remaining pole, which would be at either $\nu'=\nu$ or $\nu'=\nu_a$, labeled here as $\nu_p$ for generality, the remaining integrand becomes
    \begin{equation}
        (\pm)_p\frac{1}{2\nu_p}\frac{1}{(\nu'-\nu_p)}\sum_{i=0}^{N_{II}}\bigg(\alpha_i\prod_{\substack{k=0\\k\neq{}i}}^{N_{II}}\frac{\nu'-\nu_k}{\nu_i-\nu_k}\bigg),
    \end{equation}
    where $(\pm)_p$ just defines the sign of the term depending on whether the pole is at $\nu$ or $\nu_a$. Upon commuting sums and factors around, and comparing the integral this represents to the desired weighted sum, we see that the part of the weights $w_i$ created from the pole at $\nu_p$ is 
    \begin{equation}
        (w_i)_p=(\pm)_p\frac{1}{2\nu_p}\bigg(\prod_{\substack{k=0\\k\neq{}i}}^{N_{II}}\frac{1}{\nu_i-\nu_k}\bigg)\int_{\nu_0}^{\nu_{N_{II}}}\bigg(\prod_{\substack{k=0\\k\neq{}i}}^{N_{II}}(\nu'-\nu_k)\bigg)\frac{1}{\nu'-\nu_p} d\nu'.
    \end{equation}
    Since the current integrand is factored, we can guarantee that the pole will cancel out creating only a removable discontinuity, so long as $i\neq{}p$. In this majority of cases, the integral can be expanded and evaluated directly, since it is just a polynomial. However, if $i=p$ then the polynomial will need to be long divided to separate out the quotient of the division as a polynomial (again, easily integrated), and the portion which did not fully divide: the remainder. The remainder's contribution can be solved analytically, since it will take the form of a constant divided by the pole, where applying the Cauchy Principal Value, we find that
    \begin{equation}
        \int_{\nu_0}^{\nu_{N_{II}}}\frac{A}{\nu'-\nu_p} d\nu' = \lim_{b\rightarrow\nu_p} \log|\nu'-\nu_p| \Big|_{\nu_0}^b + \lim_{b\rightarrow\nu_p} \log|\nu'-\nu_p| \Big|_b^{\nu_{N_{II}}} = \log\bigg|\frac{\nu_{N_{II}}-\nu_p}{\nu_0-\nu_p}\bigg|.
    \end{equation}
    So long as the pole does not occur at either of the ends of the region, these two scenarios will always work. This will hold true regardless of stretching, which will further remove the points from the ends of the integration domain. To ensure this is the case, no fewer than three points may be present within any Case II region.
    \begin{figure}[H]
        \centering
        \includegraphics[width=0.7\textwidth]{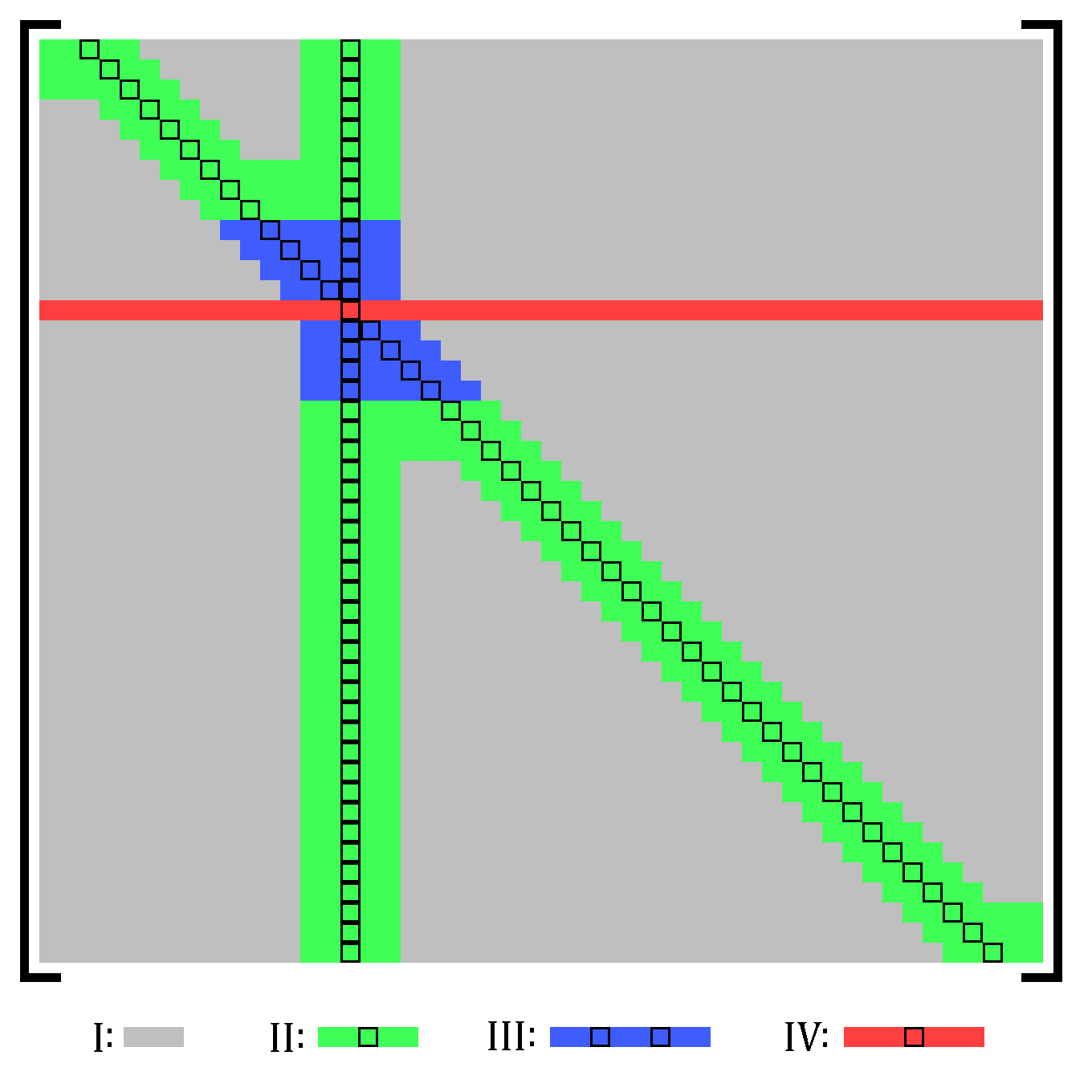}
        \caption{A graphical representation of the weights matrix $w$, where the value for $w_i^{(j)}$ is the (j,i)$^{th}$ element in the matrix. The different cases are labeled below the matrix. The black boxes designate the locations of the poles for a given iteration (row of matrix). The vertical line of poles is the fixed pole at $\nu'=\nu_a$, and the diagonal line of poles is the pole which scans across the data at $\nu'=\nu$. Where the two lines get close exemplifies when the more nuanced integration methods are needed. Regions which are still classified by Case I but are too close to a region of Case II or an end of the available data, the neighboring Case II domains are simply increased so as to avoid needing to use a lower order NC approximation due to low number of data points.}
        \label{matrix}
    \end{figure}
    
    A very similar approach may be taken for the Case III regions. Since two \em{}distinct\em{} poles are present in the region, if $\alpha$ is again represented with a Lagrange interpolating polynomial and $S$ is expanded again, there will only be one pole in each integral and the results reduce down to those of Case II. The two poles at $\nu'=-\nu$ or $\nu'=-\nu_a$ can again be treated in the Case I way. The only reason for separating this out into a new case is mostly to account for the more dynamic number of nodes included, and so it is more straightforward to place neighboring regions consistently. This variability in included nodes can be seen in the blue section of Figure \ref{matrix}, where each of these regions contains anywhere from $N_{II,min}+2$ to $2N_{II,min}+1$ points.
    
    Finally, Case IV is even simpler. If the two poles are at the same location, that means that $\nu=\nu_a$. Looking at the original definition for the integral transform, this causes $S$ to identically go to zero, making the entire integral go to zero similarly. Accordingly, there is no need to worry about different regions and arranging them for this row of the $w_i^{(j)}$ matrix. This can be seen in Figure \ref{matrix} as the single red row.

\section{Error Analysis}
    As described by the derivation shown by Isaacson and Keller\cite{isaacson}, the truncation error, $R_M$ created by replacing $\alpha$ with an $M^{th}$ degree Lagrange interpolating polynomial (with $M+1$ points) is \cite{isaacson}
    \begin{equation}
        R_M(\nu') = \bigg(\prod_{j=0}^{j=M}(\nu'-\nu_j)\bigg)\alpha[\nu_0,\nu_1,\nu_2,...,\nu_M,\nu'],
    \end{equation}
    where the latter term is the divided difference of $\alpha$. Since $\alpha$ is a well behaved function, we may assume the continuity of this divided difference regardless of $M$. Correspondingly, the error $E_M$ created by this in the standard Newton Cotes scheme is the integral of this truncation error,
    \begin{equation}
        E_M = \int_{\nu_0}^{\nu_M} \bigg(\prod_{j=0}^{j=M}(\nu'-\nu_j)\bigg)\alpha[\nu_0,\nu_1,\nu_2,...,\nu_M,\nu'] d\nu'.
    \end{equation}
    Upon making a substitution into the integral of $\nu' = \nu_0 + h t$, where $h$ is the spacing between the nodes, Isaacson and Keller work out a clean closed-form solution for the error for the standard NC scheme as
    \begin{equation}
            E_M = \frac{K_M}{(M+2)!}h^{M+3} \alpha^{(M+2)}(\xi) = \mathcal{O}(h^{M+3})
    \end{equation}
    for even $M$, and 
    \begin{equation}
        E_M = \frac{K'_M}{(M+1)!}h^{M+2} \alpha^{(M+1)}(\xi)  = \mathcal{O}(h^{M+2})
    \end{equation}
    for odd $M$. In both of these, $\xi$ is some value between $\nu_0$ and $\nu_M$. This statement is so concise due to the mean value theorem. The fact that the even and odd $M$ values have different relative accuracy, with respect to order of $h$, is from the convenient structure available inside the integrand, which those authors were able to exploit.
    
    This covers Case I, but we now need to characterize the error associated with the pole integrals investigated in this report. For Case II, there is only one pole inside the domain, so the truncation error of this pole's integral will take the form
    \begin{equation}
        R_M(\nu') = \bigg(\prod_{\substack{j=0\\j\neq{}p}}^{j=M}(\nu'-\nu_j)\bigg)\alpha[\nu_0,\nu_1,\nu_2,...,\nu_M,\nu'],
    \end{equation}
    and therefore the error associated with this 
    \begin{equation}
        E_M = \int_{\nu_0}^{\nu_M} \bigg(\prod_{\substack{j=0\\j\neq{}p}}^{j=M}(\nu'-\nu_j)\bigg)\alpha[\nu_0,\nu_1,\nu_2,...,\nu_M,\nu'] d\nu',
    \end{equation}
    and upon using that same substitution again, $\nu' = \nu_0 + h t$, we have
    \begin{equation}
        E_M = h^{M+1}\int_{0}^{M} \bigg(\prod_{\substack{j=0\\j\neq{}p}}^{j=M}(t-j)\bigg)\alpha[...,\nu_0+ht] dt = \mathcal{O}(h^{M+1}).
    \end{equation}
    Doing the same for the two distinct roots of Case III through the same procedures, we have 
    \begin{equation}
        E_M = h^{M}\int_{0}^{M} \bigg(\prod_{\substack{j=0\\j\neq{}p_1\\j\neq{}p_2}}^{j=M}(t-j)\bigg)\alpha[...,\nu_0+ht] dt = \mathcal{O}(h^{M}).
    \end{equation}
    The convenient pattern present in the default NC scheme allowed for the improvement of the odd $M$ case, but for the rest of these, that pattern is lost due to the division by the respective poles, and therefore no simplification would be straightforward or easy to find. For this same reason, it would also be difficult to apply the mean value theorem to get a concise bound for the error in terms of $\xi$. However, this is not an issue, since the scale factors (as functions of $M$) present in these error terms are necessarily finite, since they are integration of polynomials of finite degree scaled by a well behaved function ($\alpha$), evaluated over finite bounds. Accordingly, these error bounds are indeed the listed orders in $h$.
    
    If we were to consider Case IV in this manner, we would arrive at a potential issue where the integrand is not finite everywhere, since only degree of the double pole would cancel. Without further thought, this would present a serious issue with respect to bounding the error with the mean value theorem. However, this approach is not needed, since this case identically goes to zero, so the error is necessarily zero.
    
    Since the minimum number of points present in each case has been described as having a minimum, we can list out a worst case bound for the truncation error with respect to $h$ for each of the regions, as shown in Table \ref{errorOrder}. The error mentioned for each case only really describes the part of the integral which was not evaluated with a simpler case's integration scheme. 

    \begin{table}[H]
    \centering
    \caption{The orders of the errors inherent in the different types of integration used. The worst case error is described for the minimum size of each region, so that when a region is stretched, the truncation error only shrinks, and therefore is still within the listed order.}
    \label{errorOrder}
    \begin{tabular}{r|l|l}
    Case & $M_{min}$    & $E_{worst}$                 \\ \hline
    I    & $N_{I}$ even & $\mathcal{O}(h^{N_{I}+3})$  \\
    I    & $N_{I}$ odd  & $\mathcal{O}(h^{N_{I}+2})$  \\
    II   & $N_{II}$     & $\mathcal{O}(h^{N_{II}+1})$ \\
    III  & $N_{II}+1$   & $\mathcal{O}(h^{N_{II}+1})$ \\
    IV   & N/A            & 0                          
    \end{tabular}
    \end{table}

\section{Implementation}
\begin{figure}[H]
        \centering
        \includegraphics[width=1.0\textwidth]{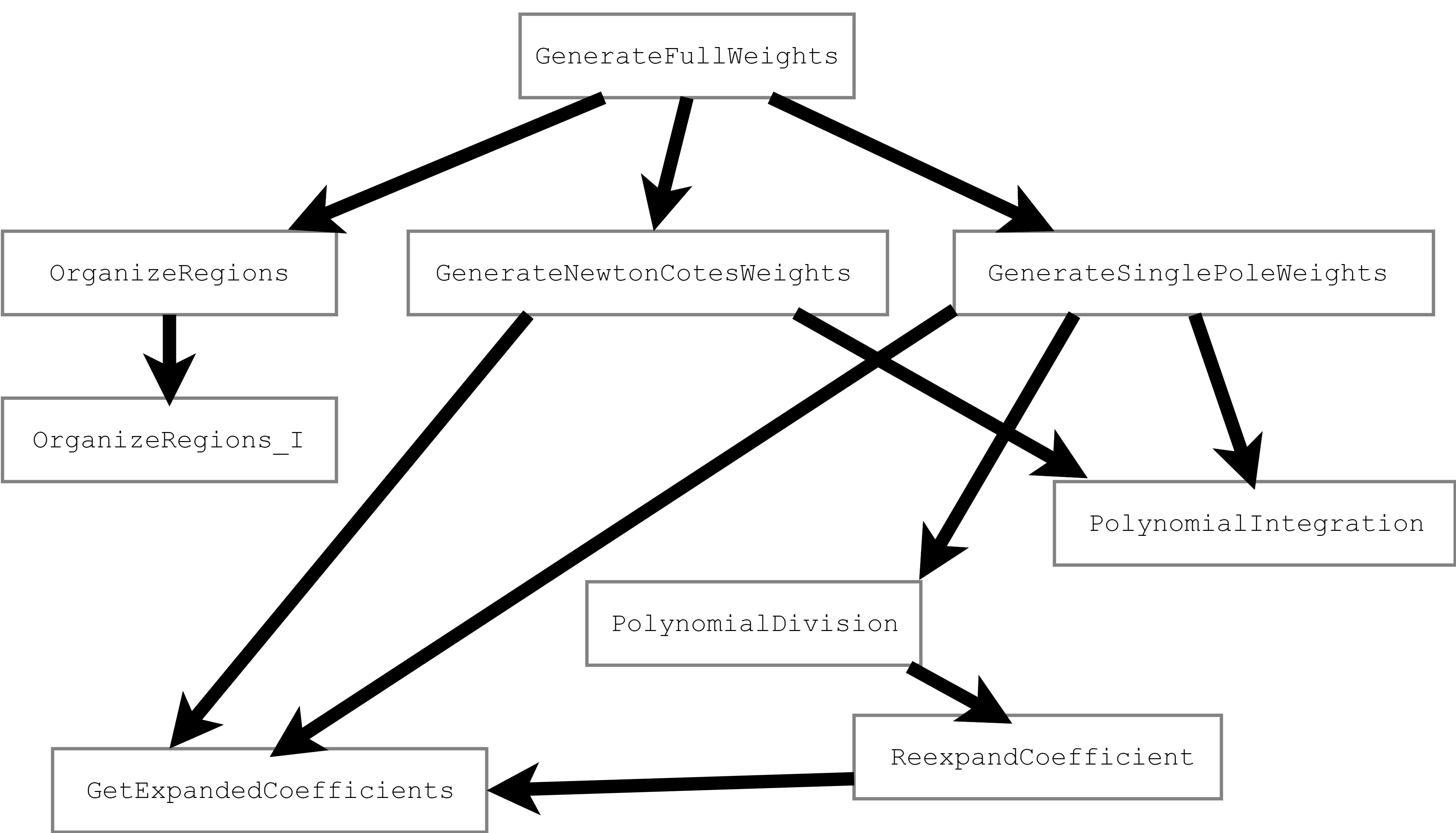}
        \caption{An outline of the \Matlab{} code implementing the procedures discussed. The arrows shown indicate their dependencies on the other functions created for this purpose.}
        \label{outline}
    \end{figure}

    These procedures were implemented inside of \Matlab{}, using $N_{I} + 1 = 3$ and $N_{II} + 1 = 5$, which ensures that the maximum error for this implementation is $\mathcal{O}(h^{5})$. Anything much higher would begin to be unreasonable, and anything too much lower would start having issues. An outline for the different functions made for this, and how they interact and depend on one another is depicted in a flow chart in Figure \ref{outline}.
    The source code for all of these files may be found in Appendices A-I, with an additional Appendix J which acts as a script/wrapper to call the functions properly, assuming the right data is available to process. Table \ref{append} details which files are in which appendix.
    \begin{table}[H]
        \centering
        \caption{My caption}
        \label{append}
        \begin{tabular}{r|l}
        Appendix & Function/Script              \\ \hline
        A        & GenerateFullWeights()        \\
        B        & OrganizeRegions()            \\
        C        & OrganizeRegions\_I()         \\
        D        & GenerateNewtonCotesWeights() \\
        E        & GenerateSinglePoleWeights()  \\
        F        & PolynomialIntegration()      \\
        G        & PolynomialDivision()         \\
        H        & ReExpandCoefficient()        \\
        I        & GetExpandedCoefficients()    \\
        J        & Test\_Script.m              
        \end{tabular}
    \end{table}
    The description of what each file does can be found in Table \ref{desc}. 
    \begin{table}[H]
        \centering
        \caption{My caption}
        \label{desc}
        \begin{tabular}{l|l}
        Function/Script              & Purpose                                                                     \\ \hline
        GenerateFullWeights()        & Generates the weight matrix to multiply $\alpha$                            \\
        OrganizeRegions()            & Handles breaking the elements of $\alpha$ into regions and cases     \\
        OrganizeRegions\_I()         & Handles the placement and stretching of Case I regions                      \\
        GenerateNewtonCotesWeights() & Generates the Case I NC weights for any number of points           \\
        GenerateSinglePoleWeights()  & Generates the Case II weights for any number of points \\
        PolynomialIntegration()      & Performs integration on arbitrary degree polynomials                        \\
        PolynomialDivision()         & Long divides a polynomial by a specified root           \\
        ReExpandCoefficient()        & Converts polynomials of $x^n$ into a different basis $(x-a)^n$              \\
        GetExpandedCoefficients()    & Generates a vector representation of a polynomial                           \\
        Test\_Script.m               & A script to demonstrate how the functions should be used                  
        \end{tabular}
    \end{table}

\subsection{Notes on Implementation}
    In addition to the lower level approach needed to actually implement the theoretical formulas into code, there were a couple minor details which might not be self-explanatory.
    
    Firstly, there is the major use of a certain scheme of polynomial manipulation, including expanding, long dividing, and integrating. These are simply writing a given polynomial in a certain basis, often $(x)^n$ where $n$ may take any integer value from zero upwards. This vector representation made for a relatively fast means for computing things like large product of first degree terms. This methodology is extended straightforwardly to integrating, since the place in the vector simply indicates what the factor becomes, and where it gets shifted to. The general scheme for writing a polynomial is
    \begin{equation}
        a_0 + a_1x + a_2x^2 + a_3x^3 + ... = [a_0, a_1, a_2, a_3, ...].
    \end{equation}
    A basis centered at $x=a$ would take the form
    \begin{equation}
        b_0 + b_1(x-a) + b_2(x-a)^2 + b_3(x-a)^3 + ... = [b_0, b_1, b_2, b_3, ...].
    \end{equation}
    From this notation it follows that multiplication by some power of $(x-a)$ in a basis centered at $a$ acts as essentially a bit shift to the right of corresponding amount. Therefore, it follows that multiplication of one vector by another could be rewritten as a sum of simple vector-times-power-of-x terms, which is straightforward from that perspective. The rest of the functionality should be similarly straightforward. 
    
    One more interesting feature is the methods by which the long division is done. A given polynomial is rewritten in the same basis as the divisor term, so that the division becomes trivial and is just a bit shift to the left, keeping the constant term off as the remainder. After the shift, the remaining polynomial can be returned to its original basis and yielded as the quotient of the division.
    
    On a different note, the two functions which are frequently called upon to generate the Case I and Case II weights are built with persistent variables. In \Matlab{} this allows the data stored in a variable to persist to future calls of that function. This functionality is used to not regenerate data if it is already stored, which just requires that the same spatial parameters are held constant between a given call. This drastically reduces the run-time of the algorithm, because it doesn't need to recalculate these smaller weights as frequently.
    
    One last detail regards how the Case II and Case III regions were assumed to be oriented within their minimum-sized sizes. Although this doesn't matter given the minimum sizes used here are both odd, if they were changed to different values then it is useful to note that a left-sided centering is used. This means that if a minimum-sized region spans an even number of points, then it will be placed on the left of the two spots in the center, instead of the right one. Accordingly, preference is given to stretch to the left before stretching to the right in order to compensate slightly. This could have been done more generally, by choosing which side offered the fewer number of stretched regions, but that would have required a much uglier case analysis within the code.

\section{Results and Discussion}
\begin{figure}[H]
        \centering
        \includegraphics[width=0.6\textwidth]{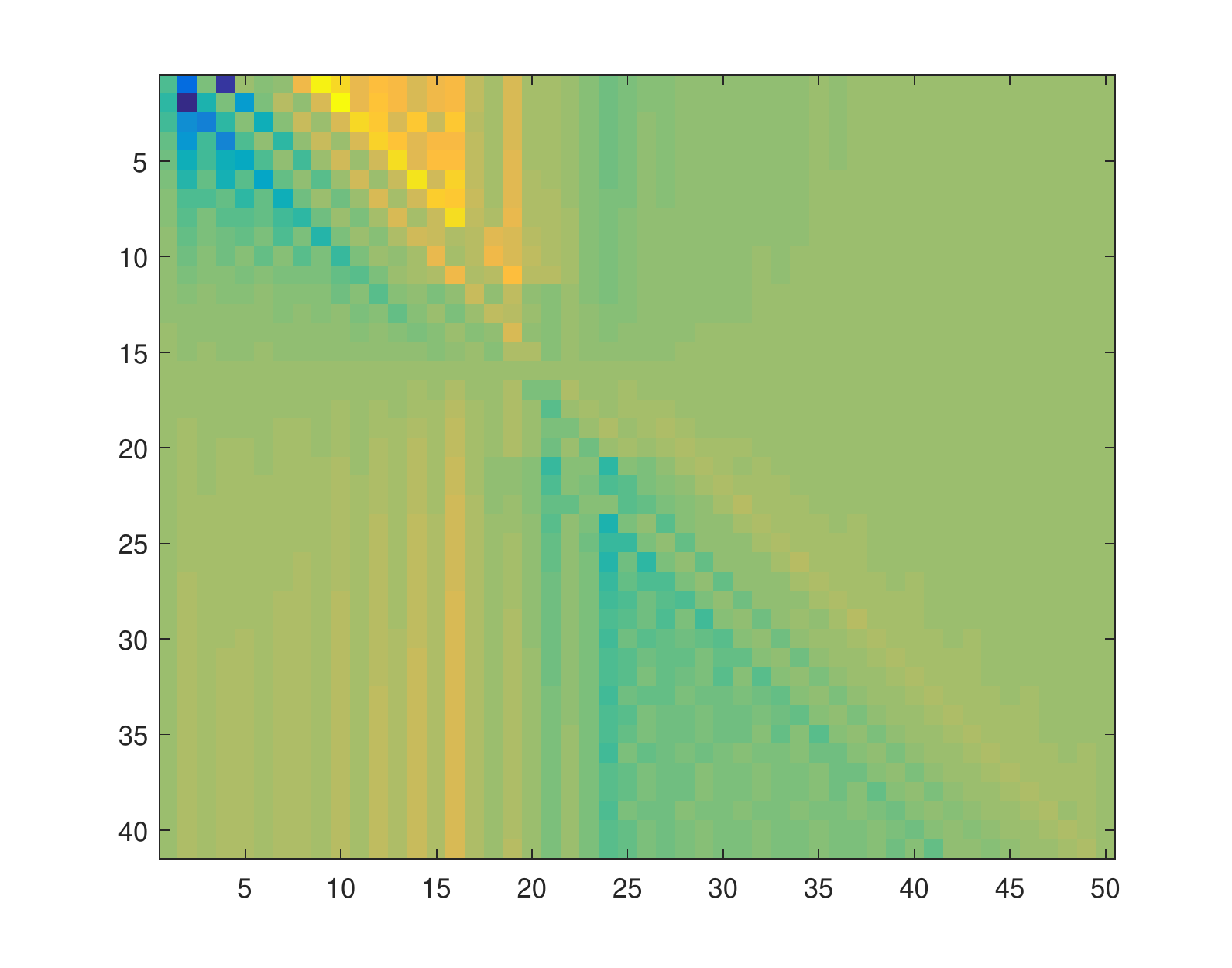}
        \caption{A visualization of the weight matrix for a small set of alpha values. This smaller set was chosen to make the patterns more visible.}
        \label{50}
    \end{figure}

\begin{figure}[H]
        \centering
        \includegraphics[width=0.6\textwidth]{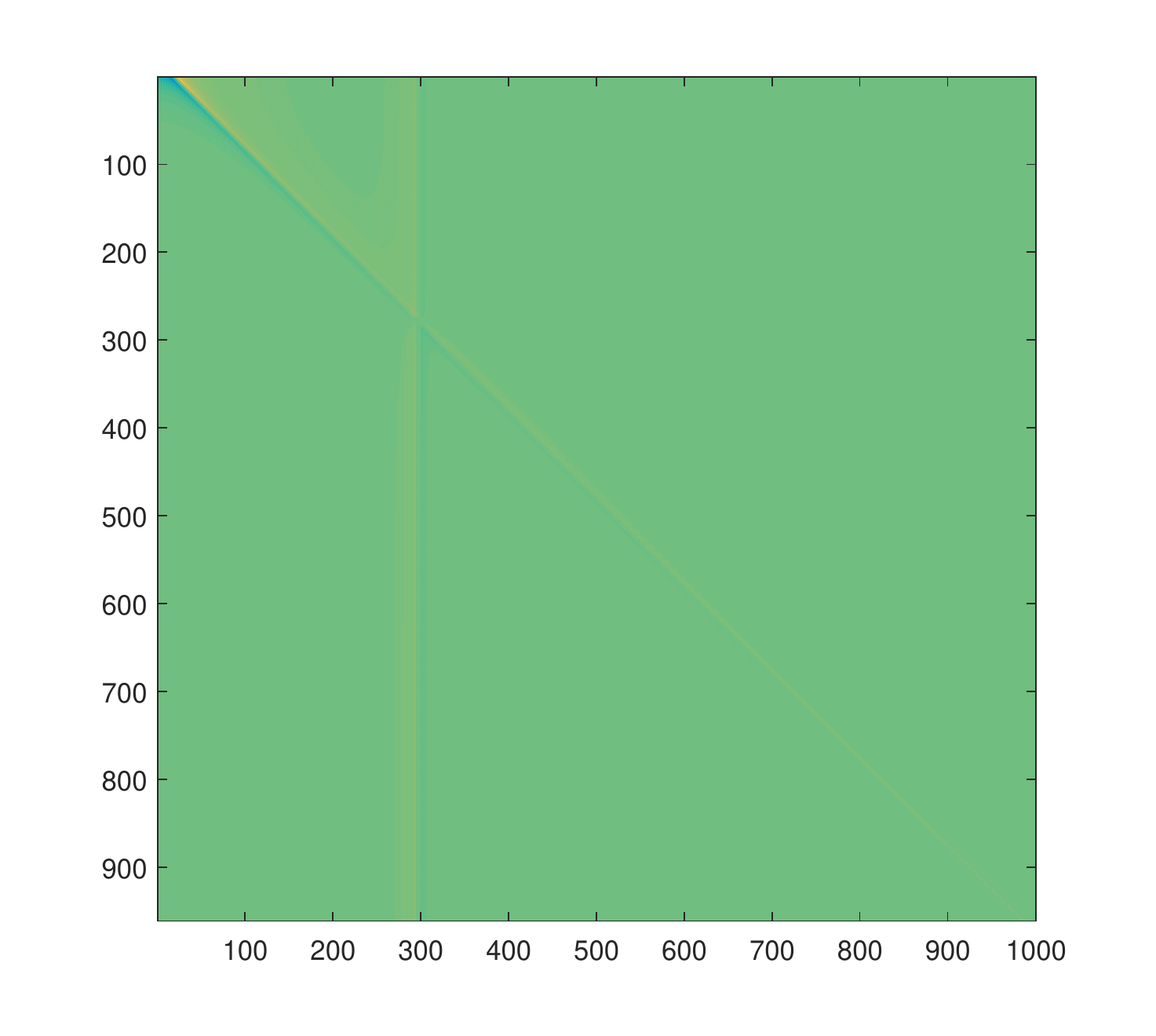}
        \caption{A visualization of the weight matrix for a larger set of alpha values. This perspective allows for easier appreciation of how localized the integration really is, since the poles are really the only visible contributors.}
        \label{1000}
    \end{figure}

    Two examples of the weight matrix generated by the appended \Matlab{} code are shown in Figures \ref{50} and \ref{1000}. The strongest features are, as expected, highly concentrated on the locations of the poles. 
    
\begin{figure}[H]
        \centering
        \includegraphics[width=0.6\textwidth]{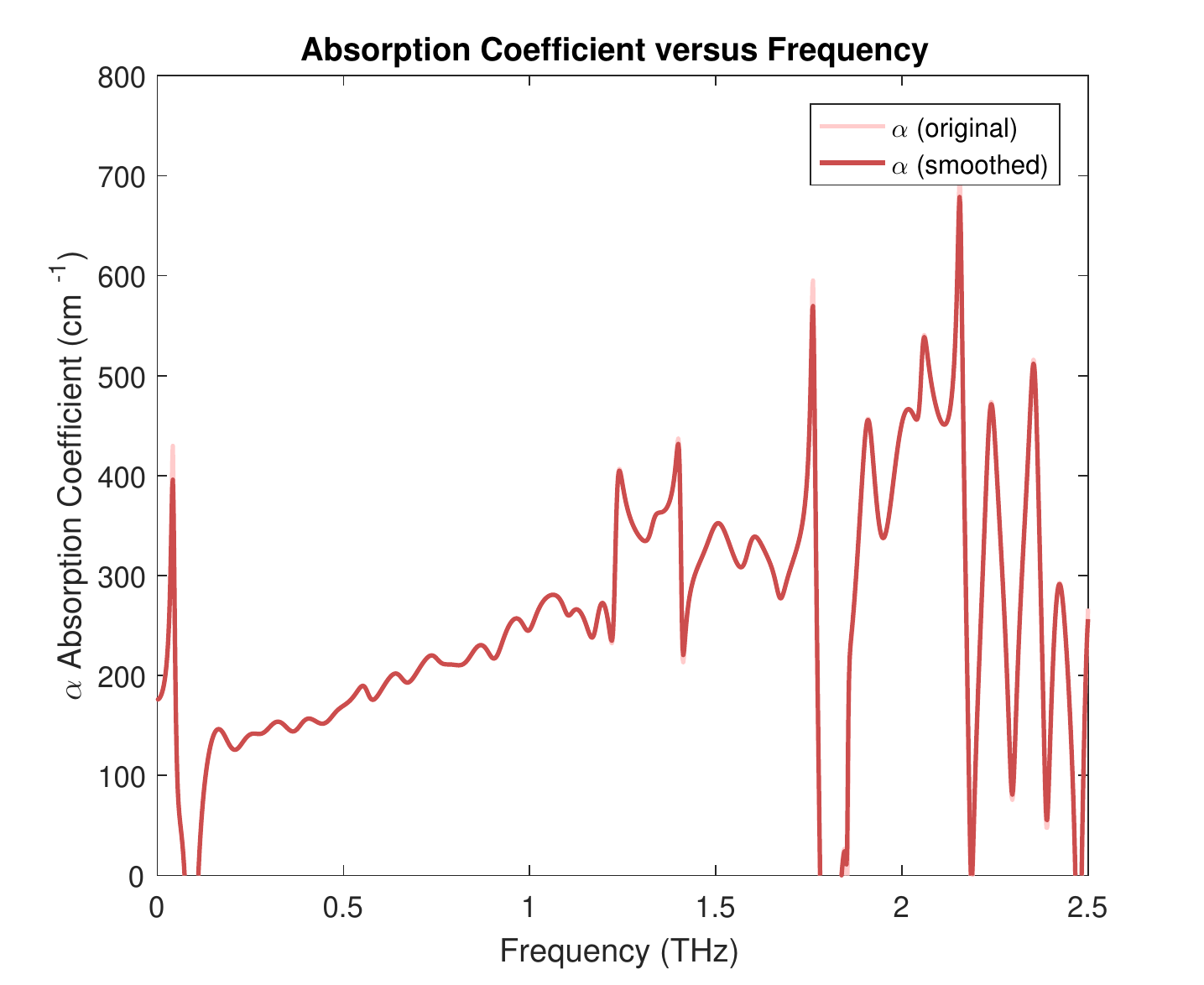}
        \caption{A sample of data for the absorption coefficient $\alpha$.}
        \label{alp}
    \end{figure}

\begin{figure}[H]
        \centering
        \includegraphics[width=0.6\textwidth]{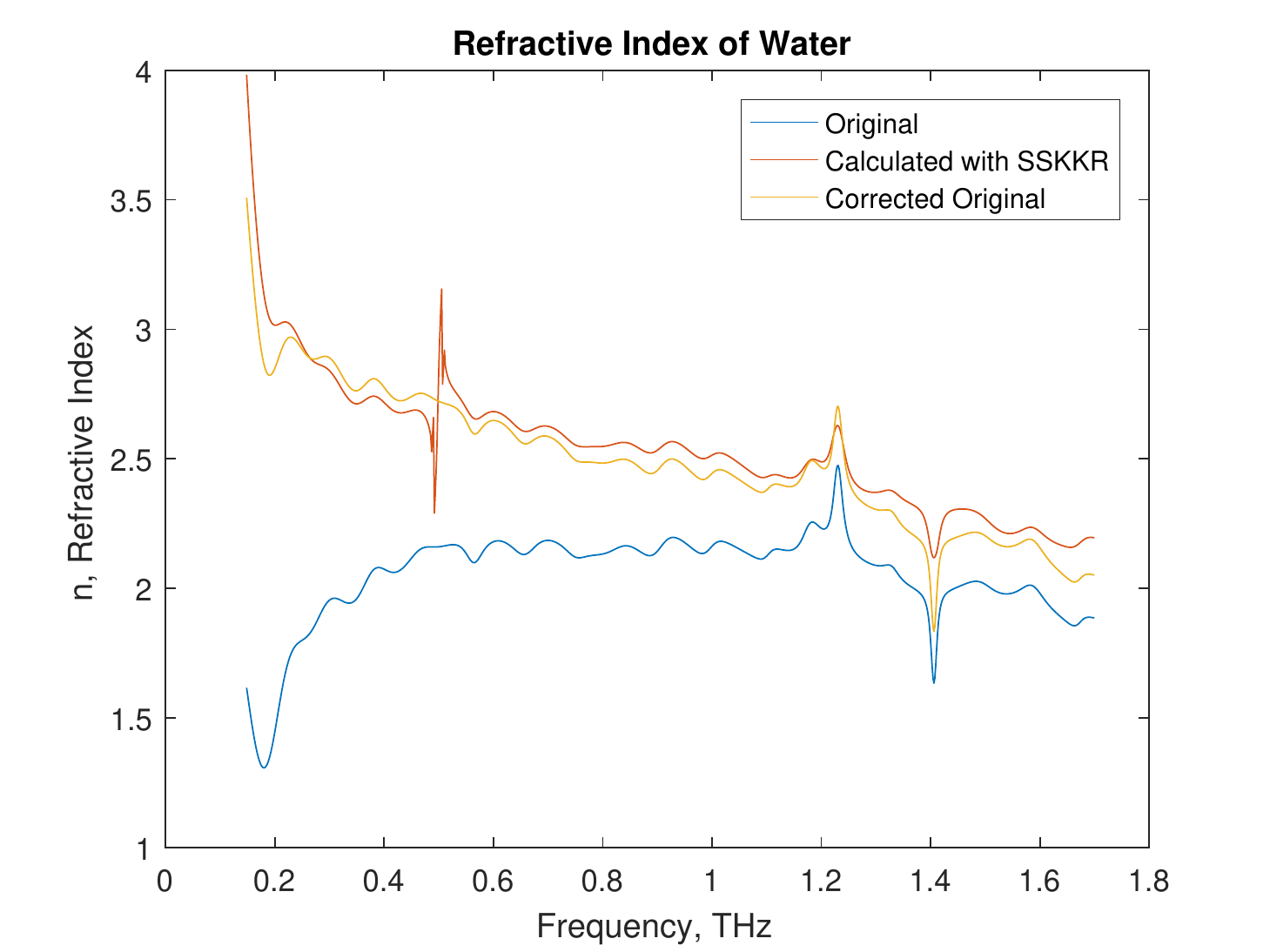}
        \caption{The refractive index corresponding to the $\alpha$, compared to the original refractive index and an externally corrected refractive index. Note that before any correction, anchor frequency would lock the resultant refractive index to the original refractive index at that point--however, here the corrected value was used as the reference instead for more direct appreciation of how the integration did. Recall that the reference only contributes an overall vertical shift, so no significant change in the form of $n$ was made under this modified display.}
        \label{refra}
    \end{figure}

    Using some actual data ($\alpha$ in Figure \ref{alp}) with the weight matrix shown in Figure \ref{1000}, we use the script file (Appendix J) to calculate and display the corresponding refractive index $n$, which is displayed in Figure \ref{refra}.\footnote{It is not recommended to use this data as a reference for the absorption and refractive index of water. While this data is accurate to itself, it us not necessarily absolutely physically accurate.}

    From these figures, it is apparent that the integral transform was successful in pulling out the refractive index, but there are significant features around the anchor frequency which are not satisfactory. It would appear the points which would be heavily influenced by Case III scenarios (when the two poles are very close) have a great deal of noise in that neighborhood. This could be a simple issue of a typo in the code, or it could be more significant, pointing to a systematic problem with the approach used in evaluating those regions.

    Upon further investigation, it becomes clear that the single pole integral results in quantities which do not depend on $h$. This makes sense from the formulas derived, but as far as an integration technique, seems quite bizarre. The next steps for this code need to be to re-examine the implementation for mistakes, and upon not finding any, re-examine how the result can be good when the points contributing the most to the integral are not affected by their spacing (and therefore their quantity). 

\section{Conclusions}
    In this report, the singly-subtractive Kramers-Kronig relations were prepared for general use in extracting information from the absorption measurements to ascertain a better understanding of the refractive index, which otherwise has a significant degree of indeterminacy. A theoretical groundwork was laid for the different scenarios that would be encountered when approximating the absorption coefficient with Lagrange interpolating polynomials, and schemes were developed for evaluating the integrals in these scenarios with heightened care taken around the iteratively moving poles of the integral transform. An implementation was discussed briefly, with the source code accompanying the report, and qualitative results confirming its functionality were presented. Features were discovered which seem counter intuitive for results of integration, but they should really be a lot worse if they are actually a problem. Further investigation will follow, digging into how significant of a problem this observed noise could be.



\section*{Supplementary material}

# Appendix A

```matlab
function weight = GenerateFullWeights(nu,alpha,anchorIndex,minIndex,maxIndex)
%Generates the full weight matrix to represent the complete integration of
%the Kramers-Kronig integral transform.
    % Inputs:
    %           nu                  = The (ordinary) frequency set corresponding to
    %                                 the set alpha. Must be equispaced ordered
    %                                 points, and should (but doesn't need to)
    %                                 start at zero frequency.
    %           alpha               = The absorption set, whose values correspond
    %                                 to whatever frequencies are present in the
    %                                 frequency set nu.
    %           anchorIndex         = The index corresponding to the anchor
    %                                 frequency in nu, which serves as a fixed
    %                                 reference point. Should be greater than
    %                                 minIndex and less than maxIndex, both by at
    %                                 least 10, if not more.
    %           minIndex            = The index of the minimum frequency in nu
    %                                 which marks the beginning of the indices for
    %                                 which the refractive index n will be
    %                                 calculated. Should be greater than 1, by at
    %                                 least 10, if not more.
    %           maxIndex            = The index of the maximum frequency in nu
    %                                 which marks the end of the indices for which
    %                                 the refractive index n will be calculated.
    %                                 Should be less than the highest possible
    %                                 index in nu, by at least 10, if not more.
    % Outputs:
    %           weight              = A matrix containing the weights which
    %                                 approximate the integral. Usage:
    %                                     n_new = n(anchorIndex) +...
    %                                         c/(2*pi^2) * (weight*alpha(:));
    %                                 where n_new will correspond to the frequencies
    %                                     nu(minIndex:maxIndex)

    %Ensure that nu and alpha are column vectors
    nu = nu(:);
    alpha = alpha(:);

    %Determine the sizes of the weight matrix
    M = length(alpha);
    N = maxIndex - minIndex + 1;
    %Determine the spacing of the equispaced points
    h = nu(2)-nu(1);

    %Define the functions for the function inside the integrand which
    %contributes to the weight of each point. Make available the full
    %function for the regions in Case I, and for the other regions: the
    %part of the function which only has poles in the negative frequency
    %region; the part which only has a pole at the scanning index; and the
    %part which only has a pole at the anchor index.
    Sfull = @(nu_,nu,nu_a) (nu^2-nu_a^2)./((nu_.^2-nu^2).*(nu_.^2-nu_a^2));
    Shalf = @(nu_,nu,nu_a) -1/(2*nu)./(nu_+nu)+1/(2*nu_a)./(nu_+nu_a);
    Snu   = @(nu_,nu)       1/(2*nu)./(nu_-nu);
```

```matlab
            Snu_a = @(nu_,nu_a)    -1/(2*nu_a)./(nu_-nu_a);

    %Initialize the weight matrix to zeros so values may easily be added to
    %their corresponding cells
    weight = zeros(N,M);

    %Cycle over each row
    for row = 1:N
            %For each row, select the location of the scanning index
            scanningIndex = row - 1 + minIndex;
            %Get a list of the different regions, and the poles present within
            %each
            [groups,localPoles] = OrganizeRegions(M,[anchorIndex,scanningIndex]);
            %Iterate over the different regions
            numGroups = length(groups);
            for g = 1:numGroups
                    %Withdraw a single group at a time, and corresponding list of
                    %poles
                    group = groups{g};
                    poles = localPoles{g};
                    %And extract the upper and lower bound for that region
                    a = group(1);
                    b = group(2);

                    %Do a bunch of case analysis:

                    %If there are no poles, treat as Case I
                    if length(poles) == 0
                            %Generate the standard NC weights
                            NC = GenerateNewtonCotesWeights(b-a,h);
                            %Scale the NC weights by the full scale function S. Store
                            %in subWeights
                            subWeights = NC .* full(nu(a:b),nu(scanningIndex),nu(anchorIndex));
                    %Only one pole exists within the region, treat as Case II
                    elseif length(poles) == 1
                            %Generate the NC weights
                            NC = GenerateNewtonCotesWeights(b-a,h);
                            %And use those NC weights for the half of the poles which
                            %are definitely no within the region (poles at negative freq).
                            subWeights = NC .* half(nu(a:b),nu(scanningIndex),nu(anchorIndex));
                            %If the pole is at the scanning frequency
                            if poles(1) + a == scanningIndex
                                    %Treat the other pole with Case I, and treat the
                                    %scanning index pole as Case II. Add the resultant
                                    %weights to the previously calculated weights
                                    subWeights = subWeights + NC .*...
                                            Snu_a(nu(a:b),nu(anchorIndex)) +...
                                            1/(2*nu(scanningIndex))*...
                                            GenerateSinglePoleWeights(b-a,h,scanningIndex-a);
                            %If the pole is instead at the anchor frequency
                            else
                                    %Treat the other pole with Case I, and treat the anchor
                                    %index pole as Case II. Add the resultant weights to
                                    %the previously calculated weights
                                    subWeights = subWeights + NC .*...
                                            Snu(nu(a:b),nu(scanningIndex)) -...
                                            1/(2*nu(anchorIndex))*...
```

```matlab
                                        GeneratePoleWeights(b-a,h,anchorIndex-a);
                            end
                        %If there are two poles within the region (Could be Case III or IV)
                        else
                            %If the poles are different, treat as Case III
                            if poles(1) ~= poles(2)
                                %Generate the NC weights for this region
                                NC = GenerateNewtonCotesWeights(b-a,h);
                                %Calculate the weights which correspond to the poles at
                                %negative frequency
                                subWeights = NC .* ...
                                    Shalf(nu(a:b),nu(scanningIndex),nu(anchorIndex));
                                %And add to it the Case I weights from each pole
                                subWeights = subWeights + ...
                                    1/(2*nu(scanningIndex))*...
                                    GenerateSinglePoleWeights(b-a,h,scanningIndex-a)...
                                    -1/(2*nu(anchorIndex))*...
                                    GenerateSinglePoleWeights(b-a,h,anchorIndex-a);
                            %If the poles are equal, treat as Case IV
                            else
                                %Set the weights to zero, since the integral is
                                %identically zero in this case. This region will
                                %contain all of that row, by design of the
                                %OrganizeRegions() function.
                                subWeights = zeros(b-a+1,1);
                            end
                        end

                        %Gather the results of the case analysis and add the results to
                        %the full weight matrix in the correct locations
                        weight(row,a:b) = weight(row,a:b) + reshape(subWeights,1,[]);
                end
            end
        %Return the full weight matrix
    end
```

# Appendix B

```matlab
function [ groups,localPoles ] = OrganizeRegions(numAlpha,poleLocations)
%Breaks the row of alpha indices into regions to be solved with different
%cases, and who have different pole proximity. This function handles
%stretching regions so they fit better.
    % Inputs:
    %           numAlpha            = The number of alpha values to be broken into
    %                                   groups/regions
    %           poleLocations   = A vector containing the locations of the
    %                                   poles present for the given row
    % Outputs:
    %           groups              = A cell array row vector containing the index
    %                                   of the first and last alpha index in the
    %                                   group. The entries in this cell array
    %                                   correspond directly to the entries in the
    %                                   localPoles
    %           localPoles          = A cell array row vector containing the
    %                                   relative index values of the poles present
    %                                   within each group. These are relative to the
    %                                   left end of the respective group. Each entry
    %                                   can have no poles, or 1 pole, or two poles.
    %                                   A double pole present in a region simply has
    %                                   the same index repeated in the list for that
    %                                   group

    %The minimum number of points to be contained within the Case I regions
    minI = 3;
    %The minimum number of points to be contained within the Case II
    %regions
    minII = 5;

    %Extract the location of the two poles
    pole1 = min(poleLocations);
    pole2 = max(poleLocations);

    %If the poles are equal, treat as case IV
    if pole1==pole2
        %Make one big group containing the entire row.
        groups = {[1,numAlpha]};
        %Propagate the values for the poles to list both of them
        localPoles = {poleLocations};
    %Now handle the other cases
    else
        %Initialize groups and localPoles as cell arrays
        groups = {};
        localPoles = {};
        %Handle the region to the left of the leftmost Case II region:
        %If that region is large enough to fit at least one of the Case I
        %regions, then place as many as possible, prioritizing lower size.
        %Stretch as needed.
        if pole1 - ceil(minII/2) >= minI
            %Determine the left edge of the leftmost pole's region
            pole1Left = pole1 - ceil(minII/2) + 1;
            %Generate the subgroups present in the region to the left of
```

```matlab
            %that edge, and have stretching prioritize on the right
            subGroups = OrganizeRegions_I(1,pole1Left-1,minI,'right');
            %Determine the number of subgroups generated
            num1 = length(subGroups);
            %Append these subgroups to the group cell array
            groups(end+(1:num1)) = subGroups;
            %Append empty vectors to the localPoles cell array, since these
            %are all Case I regions and therefore do not have poles within
            %them
            localPoles(end+(1:num1)) = {[]};
        %If the region to the left of the leftmost pole does not have
        %enough room to fit even one minimum sized Case I region, then omit
        %this Case I region, and simply stretch the Case II region over to
        %the left all the way
        else
            pole1Left = 1;
        end

        %Handle the region to the right of the rightmost Case II region:
        %If that region is large enough to fit at least one of the Case I
        %regions, then place as many as possible, prioritizing lower size.
        %Stretch as needed.
        if pole2 + floor(minII/2) + 1 <= numAlpha - minI + 1
            %Determine the right edge of the rightmost pole's region
            pole2Right = pole2 + floor(minII/2);
            %Generate the subgroups present in the region to the right of
            %that edge, and have stretching prioritize on the left
            subGroups = OrganizeRegions_I(pole2Right+1,numAlpha,minI,'left');
            %Determine the number of subgroups generated
            num5 = length(subGroups);
            %Append these subgroups to the group cell array
            groups(end+(1:num5)) = subGroups;
            %Append empty vectors to the localPoles cell array, since these
            %are all Case I regions and therefore do not have poles within
            %them
            localPoles(end+(1:num5)) = {[]};
        %If the region to the right of the rightmost pole does not have
        %enough room to fit even one minimum sized Case I region, then omit
        %this Case I region, and simply stretch the Case II region over to
        %the right all the way
        else
            pole2Right = numAlpha;
        end

        %Now tackle the indices between the left edge of the leftmost pole
        %region and the right edge of the rightmost pole region:
        %Compute the distance between the right edge of the rightmost
        %pole's region and the left edge of the rightmost pole's region.
        maxDistance = (pole2 - ceil(minII/2)) - (pole1 + floor(minII/2) + 1) + 1;
        %If the regions are so close that they overlap, treat as Case III
        if maxDistance < 0
            %Group the entire remaining set of indices into one group
            groups{end+1} = [pole1Left,pole2Right];
            %Keep both poles listed within the localPoles for this group.
            %Subtract off by the left edge of the leftmost pole's orginal
            %region so the values ouputed is relative to this lower edge.
            localPoles{end+1} = [pole1,pole2] - pole1Left;
```

```matlab
                %If there is not enough space to fit a full Case I region between
                %the two pole regions, treat as Case II, but stretch the inner
                %edges of each pole region inwards to cover the indices where a
                %Case I region would have gone.
                elseif (maxDistance < minI) && (maxDistance >= 0)
                    %Compute the stretched right edge of the leftmost pole's region
                    pole1Right = pole1 + floor(minII/2) + floor(maxDistance/2);
                    %And the same for the left edge of the rightmost pole's region
                    pole2Left  = pole2 - ceil(minII/2) + 1 - ceil(maxDistance/2);
                    %Store the leftmost pole's region in groups
                    groups{end+1} = [pole1Left,pole1Right];
                    %as well as the corresponding relative pole value
                    localPoles{end+1} = pole1-pole1Left;
                    %and do the same thing for the other rightmost pole's region
                    %too
                    groups{end+1} = [pole2Left,pole2Right];
                    localPoles{end+1} = pole2-pole2Left;
                %If there is enough room between them to allow for at least one
                %minimum sized Case I region, then leave the poles' Case II regions
                %unstretched, and fill the in-between with Case I regions
                else
                    %Calculate the right edge of the leftmost pole's region
                    pole1Right = pole1 + floor(minII/2);
                    %and the same for the left edge of the rightmost pole's region
                    pole2Left  = pole2 - ceil(minII/2) + 1;
                    %Save the corresponding regions and single-pole values in the
                    %groups and localPoles cell arrays
                    groups{end+1} = [pole1Left,pole1Right];
                    localPoles{end+1} = pole1-pole1Left;
                    groups{end+1} = [pole2Left,pole2Right];
                    localPoles{end+1} = pole2-pole2Left;
                    %Determine the subgroups within the Case I region between the
                    %two poles' regions.
                    subGroups = OrganizeRegions_I(pole1Right+1,pole2Left-1,minI,'sides');
                    %Determine the number of subgroups
                    num3 = length(subGroups);
                    %Append these subgroups onto groups
                    groups(end+(1:num3)) = subGroups;
                    %and also append empty vectors to localPoles, since these are
                    %all Case I regions and therefore do not have poles present
                    localPoles(end+(1:num3)) = {[]};
                end
        end
        %Return the groups and localPoles cell arrays
end
```

# Appendix C

```matlab
function regions = OrganizeRegions_I(a,b,minSize,skewness)
%Generates the set of subgroups which will cover a given span of indices.
%This will first prioritize a minimum number of subgroups all with at least
%the specified number of indices within; then it prioritizes distributing
%the needed stretch amount over as many of the subgroups as possible; then
%it will prioritize the location of these subgroups based on the specified
%skewness.
    % Inputs:
    %           a           = The index of the left edge of the region to be
    %                         covered
    %           b           = The index of the right edge of the region to be
    %                         covered
    %           minSize     = The minimum number of points which may be present
    %                         in each subgroup
    %           skewness    = Where to put the most stretched subgroups within
    %                         the set. Valid options are
    %                               'left'   = largest on left
    %                               'right'  = largest on right
    %                               'sides'  = largest on the ends, smalles in
    %                                          the center
    % Outputs:
    %           regions     = A cell array, whose elements are a vector
    %                         containing the indices corresponding to the
    %                         beginning and end of the subgroup's region

    %Compute the full length of the region
    currentLength = b-a+1;
    %Determine how much stretch is needed
    stretchNeeded = mod(currentLength,minSize);
    %Store the actual number of subgroups to be placed
    maxNum = floor(currentLength/minSize);
    %Initialize a vector called sizes to store the lengths of each subgroup
    sizes = nan(1,maxNum);
    %Iterate over these subgroups and determine a list of their lengths
    for segment = 1:maxNum
        %For each, choose the smallest upper bound for how much
        %stretching is needed assuming the stretching is broken up as
        %uniformly as possible
        stretch = ceil(stretchNeeded/maxNum);
        %Define the length of the current subgroup to be the minimum size
        %plus this stretch amount
        sizes(segment) = minSize + stretch;
        %Update the amount of stretch needed
        stretchNeeded = stretchNeeded - stretch;
    end

    %Then reorder these lengths so that they have the desired skewness:
    %If left, then make no change
    if strcmp(skewness,'left')
        ordered = sizes;
    %If right, then simply reverse the list of lengths
    elseif strcmp(skewness,'right')
        ordered = fliplr(sizes);
```

```matlab
        %If sides, then
        elseif strcmp(skewness,'sides')
            %Start the ordered set with the first length (longest)
            ordered = sizes(1);
            %For the remaining lengths, repeatedly add them to the center of
            %the ordered set, so that the largest ones get pushed outwards to
            %the sides
            for i = 1:(maxNum-1)
                %Calculate the center
                pos = ceil(length(ordered)/2)+1;
                %Split, insert, and concatenate results
                ordered = [ordered(1:pos-1),sizes(i+1),ordered(pos:i)];
            end
        end
        
        %Initialize the ouput variable, regions
        regions = cell(1,maxNum);
        %Calculate the list of indices which will be needed, relative to the
        %left edge, at 'a'
        placement = [0,cumsum(ordered)-1];
        %Grab the two respective ends of each subgroup, and place it in the
        %output variable. Iterate over all of these consecutive regions
        for segment = 1:maxNum
            regions{segment} = a+[placement(segment),placement(segment+1)];
        end
        %Return the cell array of each subgroup's region
end
```

# Appendix D

```matlab
function weights_ = GenerateNewtonCotesWeights(n,h)
%Generates the standard Newton-Cotes integration weights with the specified
%order and node spacing
    % Inputs:
    %       n               = The order of the integration approximation. Equal to
    %                         the number of nodes to be used minus 1.
    %       h               = The node spacing. Nodes must be equispaced.
    % Outputs:
    %       weights_        = A vector containing the corresponding n+1 NC
    %                         weights, which approximate the integral

    %Store the value for n in a separate variable
    currentOrder = n;
    %Define persistent variables weights and order. These will allow the
    %data stored within them to persist between calls of this function.
    persistent weights order
    %If weights and order have been defined in a previous call of this
    %function, AND the order is the same as it was previously
    if isempty('weights')~=1 && isempty('order')~=1 ...
                        && isequal(order,currentOrder)
        %then don't bother recalculating and just return the previous value
        weights_ = h*weights;
    %If it was not calculated before or the order has changed, perform a
    %fresh calculation
    else
        %Initialize the weights
        weights = nan(n+1,1);
        %Iterate over each element to be calculated in weights
        for i = 0:n
            %Create the list of indices, following the Lagrange polynomial
            %formulation, which simply excludes the ith index
            jSet = [0:(i-1),((i+1):n)];
            %Find the expanded form of this large product of indices
            expCoeffs = GetExpandedCoefficients(jSet);
            %Integrate this resultant polynomial from 0 to n
            intResult = PolynomialIntegration(expCoeffs,0,n);
            %Calculate the scale factor
            scaleFactor = 1/prod(i-jSet);
            %Store the resultand full integral result
            weights(i+1) = intResult * scaleFactor;
        end
        %Store the values scaled by h in the output variable
        weights_ = h*weights;
        %Update the value for order so it may be properly used in the next
        %call
        order = currentOrder;
    end
    %Return weights_
end
```

# Appendix E

```matlab
function weights_ = GenerateSinglePoleWeights(n,h,p)
%Generates the modified Newton-Cotes integration weights with the specified
%order and node spacing, when there is a first order pole at the pth node
    % Inputs:
    %        n              = The order of the integration approximation. Equal to
    %                          the number of nodes to be used minus 1.
    %        h              = The node spacing. Nodes must be equispaced.
    %        p              = The index relative to the start of the region of
    %                          where the pole is.
    % Outputs:
    %        weights_       = A vector containing the corresponding n+1 NC
    %                          weights, which approximate the integral

    %Store the values for n and p in a separate variable
    currentOrder = n;
    currentPlacement = p;
    %Define persistent variables weights, order, and placement. These will
    %allow the data stored within them to persist between calls of this
    %function.
    persistent weights order placement
    %If weights, order, and placement have been defined in a previous call
    %of this function, AND the order and placement is the same as it was
    %previously
    if isempty('weights')~=1 && isempty('order')~=1 && isempty('placement')~=1 ...
                    && isequal(order,currentOrder)...
                    && isequal(placement,currentPlacement)
        %then don't bother recalculating and just return the previous value
        weights_ = h*weights;
    %If it was not calculated before or the order or placement has changed,
    %perform a fresh calculation
    else
        %Initialize the weights
        weights = nan(n+1,1);
        %Iterate over each element to be calculated in weights
        for i = 0:n
            %Create the list of indices, following the Lagrange polynomial
            %formulation, which simply excludes the ith index
            jSet = [0:(i-1),((i+1):n)];
            %Find the expanded form of this large product of indices
            expCoeffs = GetExpandedCoefficients(jSet);
            %Perform a long division of this polynomial by a pole at p.
            %Store the quotient polynomial which is fully divided, and
            %also the remainder of the division. The remainder should be
            %zero unless i=p
            [quotient,remainder] = PolynomialDivision(expCoeffs,p);
            %Integrate this resultant divided polynomial from 0 to n, and
            %add onto it the result of integrating the remainder function,
            %which is just a natural log solution.
            intResult = PolynomialIntegration(quotient,0,n) +...
                        remainder * log(abs((n-p)/(p)));
            %Calculate the scale factor, which was unaffected by the pole
            scaleFactor = 1/prod(i-jSet);
            %Store the resultand full integral result
```

```matlab
                    weights(i+1) = intResult * scaleFactor;
            end
            %Store the values scaled by h in the output variable
            weights_ = h*weights;
            %Update the value for order and placement it may be properly used
            %in the next call
            order = currentOrder;
            placement = currentPlacement;
        end
        %Return weights_
end
```

# Appendix F

```matlab
function intResult = PolynomialIntegration(expCoeffs,lb,ub)
%Computes the integral over a polynomial with the specified coefficients
%over the specified bounds
    % Inputs:
    %           expCoeffs   =   The coefficient vector generated by
    %                           GetExpandedCoefficients() which represents the
    %                           polynomial in the x^k basis
    %           lb          =   The lower bound, x=lb, of the integral
    %           ub          =   The upper bound, x=ub, of the integral
    % Outputs:
    %           intResult   =   The numerical value of the integral of the
    %                           specified polynomial over the specified bounds.

    %Calculate the number of coefficients present. The polynomial is
    %therefore order num-1
    num = length(expCoeffs);
    %Calculate the exponents which result from integrating the x^k terms,
    %where 0 <= k <= num-1
    exponents = (1:num);
    %Calculate the modified coefficients after dividing elementwise by the
    %exponent (mimicing standard polynomial integration)
    newCoeffs =  expCoeffs ./ exponents;

    %Define the x values for the upper and lower bounds as vectors so they
    %may more easily be raised to a power, multiplied, and then summed
    %afterwards.
    LB = repmat(lb,1,num);
    UB = repmat(ub,1,num);

    %Determine the integral, taking the difference of each term's
    %corresponding value of the antiderivative at the upper and lower
    %bounds
    intResult = sum( ( UB.^exponents - LB.^exponents ) .* newCoeffs );
    %Return intResult
end
```

# Appendix G

```matlab
function [quotient,remainder] = PolynomialDivision(expCoeffs,divRoot)
%Compute the long division of a given polynomial by a divisor root. Returns
%the main quotient and the remainder of the division.
    % Inputs:
    %           expCoeffs       = The coefficient vector generated by
    %                             GetExpandedCoefficients() which represents the
    %                             polynomial in the x^k basis
    %           divRoot         = The location of the root in the first degree
    %                             divisor.
    % Outputs:
    %           quotient        = The quotient of the polynomial division, in the
    %                             x^k basis
    %           remainder       = The remainder of the division

    %Long division is complicated, but if the polynomial were represented
    %in the basis of the divisor root, then the division would be trivial
    %and would simply be a straightforward cancellation. The zeroth degree
    %term in that basis would be the remainder, and all the higher degree
    %terms would just get bitshifted down, and after being retransformed in
    %the original basis, are the quotient.

    %Re-express the polynomial in the divisor root's basis
    reExpand = ReExpandCoefficients(expCoeffs,divRoot);
    %Extract the remainder
    remainder = reExpand(1);
    %Trim off the first element of the re-expressed polynomial and then
    %re-express it back in the original basis.
    quotient = ReExpandCoefficients(reExpand(2:end),-divRoot);

    %Return quotient and remainder
end
```

# Appendix H

```matlab
function expCoeffs_new = ReExpandCoefficients(expCoeffs,newCenter)
%Re-expressed a vector of coefficients expCoeffs which are defined in the
%x^k basis, in terms of a new basis (x-newCenter)^k. The polynomial is
%constructed to remain equivalent, but just have different apparent
%coefficients (since it's in a new basis).
    % Inputs:
    %           expCoeffs       = The coefficient vector generated by
    %                             GetExpandedCoefficients() which represents the
    %                             polynomial in the x^k basis
    %           newCenter       = The desired center of the new basis
    % Outouts:
    %           expCoeffs_new   = The coefficient vector corresponding to the
    %                             same polynomial represented in the new basis of
    %                             (x-newCenter)^k

    %Essentially this process will use the fact that
    %       (x)^k = (x+newCenter-newCenter)^k
    %             = ((x+newCenter)-newCenter)^k
    %and therefore just need to find expansions of (x+newCenter)^k since in
    %the new basis this last equation is represented by
    %             = (x+newCenter)^k

    %Determine the degree of the polynomial
    num = length(expCoeffs) - 1;
    %Initialize the output variable
    expCoeffs_new = zeros(size(expCoeffs));
    %Populate the zeroth degree entry in the output variable as the zeroth
    %degree entry in the original basis, since this is the one component
    %which is not directly transformed.
    expCoeffs_new(1) = expCoeffs(1);
    %Iterate over the different degree terms remaining, (x+newCenter)^k,
    %expand them, and add their corresponding entries to the output
    %variable
    for i = 1:num
        %Create a list of the roots, which are all the same, and have
        %degree of i
        newCenter_ = repmat(-newCenter,1,i); %Negative is since (x+...)^k not (x-...)^k
        %Expand this power, scale it by the original coefficient for that
        %degree, and add the resulttant polynomial coefficients onto the
        %outout
        expCoeffs_new = expCoeffs_new +...
            expCoeffs(i+1)*...
                padarray(GetExpandedCoefficients(newCenter_),[0,num-i],0,'post');
    end
    %Return the recentered/new-basis polynomial's coefficients
end
```

# Appendix I

```matlab
function expCoeffs = GetExpandedCoefficients(roots)
%Generate a vector representative of the expanded version of the polynomial
%with the specified roots
    % Inputs:
    %           roots       =   A vector containing the roots of the polynomial,
    %                           fully representing its factored form, i.e.
    %                                   (x-r1)(x-r2)(x-r3)...(x-rn)
    %                           where r1 through rn are the roots listed in roots.
    %                           Must have at least one root.
    % Outputs:
    %           expCoeffs   =   A vector representing the coefficients of the
    %                           polynomial in expanded form whose entries increase
    %                           in degree. Ex: the polynomial of the form
    %                                   a0 + a1*x + a2*x^2 + a3*x^3 + ...
    %                           is represented in vector form as
    %                                   [a0, a1, a2, a3, ...]
    %                           Essentially, this just represents a vector, whose
    %                           kth basis is x^k, k>=0.

    %Determine the number of roots
    num = length(roots);

    %Initialize the set of expanded coefficients with the first root,
    %written in the described form
    expCoeffs = [-roots(1),1];

    %For all the remaining terms, iteratively distribute scaling by (x-rk).
    %The x part essentially acts similarly to a bit shift operator, and the
    %-rk part scales the previous vector. Generate these recursively,
    %repeatedly summing the result. Use padarray to ensure that the vectors
    %are of the same length so Matlab can add them.
    for term = 2:num
            expCoeffs = padarray(expCoeffs,[0,1],'pre') -...
                            roots(term)*padarray(expCoeffs,[0,1],'post');
    end

    %Return expCoeffs
end
```

# Appendix J

```matlab
load('AbsorptionData.mat') %minIndex will be 90; maxIndex will be 1020

%This value should be around 300 for the given data
anchorIndex = 300;
%Define the speed of light
c = 0.299792458; %mm*THz


%Generate the weight matrix
W = GenerateFullWeights(freqSet,a1,anchorIndex,minIndex,maxIndex);

%Visualize the weight matrix
figure;
imagesc(W);
daspect([1 1 1])

%Calculate the new refractive index
n_new = n2(anchorIndex) + c / (2*pi^2) * ( W * a1(:) );

%Visualize the refractive index
figure;
%Plot the original refractive index
plot(freqSet(minIndex:maxIndex),n0(minIndex:maxIndex));
hold on;
%Plot the new refractive index
plot(freqSet(minIndex:maxIndex),n_new(:));
%Plot the corrected refractive index (corrected by carefully fitting it to
%the n_new)
plot(freqSet(minIndex:maxIndex),n2(minIndex:maxIndex));
legend({'Original','Calculated with SSKKR','Corrected Original'});
title('Refractive Index of Water');
xlabel('Frequency, THz');
ylabel('n, Refractive Index');
```



\end{document}